\documentclass[useAMS,usenatbib]{mn2e}

\usepackage{epstopdf,graphicx,amsmath,amssymb,color}
\title[Importance of thermal diffusion]{Importance of thermal diffusion in the gravo-magnetic limit cycle}
\author[Owen, J.E. \& Armitage, P.J.]{James E. Owen$^{1}$\thanks{E-mail: jowen@cita.utoronto.ca} and Philip J. Armitage$^{2,3}$\thanks{E-mail: pja@jilau1.colorado.edu}  \\
$^{1}$Canadian Institute for Theoretical Astrophysics, 60 St. George Street, Toronto, M5S 3H8, Canada.\\
$^{2}$JILA, University of Colorado and NIST, 440 UCB, Boulder, CO 80309-0440, USA.\\
$^{3}$Department of Astrophysical and Planetary Sciences, University of Colorado, Boulder, CO 80309-0391, USA}

\newcommand{\msunyr}{M$_\odot$~yr$^{-1}$}

\newcommand{\bc}{}

\begin{document}

\pagerange{\pageref{firstpage}--\pageref{lastpage}} \pubyear{2002}

\maketitle

\label{firstpage}

\begin{abstract}
We consider the role of thermal diffusion due to turbulence and radiation on accretion bursts that occur in protoplanetary discs which contain dead zones. Using 1D viscous disc models we show that diffusive radial transport of heat is important during the gravo-magnetic limit cycle, and can strongly modify the duration and frequency of accretion outbursts. When the Prandtl number is large - such that turbulent diffusion of heat is unimportant - radial radiative diffusion reduces the burst duration compared to models with no diffusive transport of heat. When the Prandtl number is small ($\lesssim 25$) we find that turbulent diffusion dominates the radial transport of heat, reducing the burst duration to $\lesssim 10^3$~years as well as increasing the burst frequency. Furthermore, inclusion of radial transport of heat extends the range of infall rates under which the disc undergoes accretion bursts from $10^{-8}$ to $10^{-6}$ \msunyr with no diffusion, to $10^{-8}$ to $\gtrsim10^{-4}$ \msunyr with radiative and strong turbulent diffusion. The relative roles of radiative and turbulent thermal diffusion are likely to vary during an accretion burst, but 
simple estimates suggest that the expected Prandtl numbers are of the order of 10 in protoplanetary discs, and hence 
that turbulent diffusion is likely to be an important process during accretion outbursts due to the gravo-magnetic limit cycle.
\end{abstract}


\section{Introduction}
A subset of protostellar systems exhibit large amplitude variations in disc luminosity that appear to 
reflect discrete accretion episodes through the inner disc and on to the star \citep{audard14}. The most dramatic known 
events are those that occur in FU Orionis systems \citep{hartmann96}, while in EXor systems both the amplitude and 
duration of accretion bursts are reduced. Given the sparsity of historical observations it is possible 
that all protostars undergo outbursting behaviour (and that 
additional classes of very long duration outbursts remain unrecognised). Indeed, statistical arguments 
based on a comparison between the observed integrated luminosity of protostars and their final mass 
{\bc \citep[the ``protostellar luminosity problem",][]{kenyon90,evans09}} are frequently taken to imply that large-scale variability is 
ubiquitous.

The reason or reasons why protostellar discs show outbursts has not been definitively established. 
One class of models attributes the outbursts to an instability in discs that contain a dead-zone \citep{gammie96}, 
which in the simplest (original) description is a radial region where the strength of vertically integrated turbulence 
is suppressed by Ohmic damping of the magnetorotational instability \citep[MRI,][]{balbus98}. Outside the thermally ionized inner 
disc, Ohmic damping can potentially reduce the {\em maximum} accretion capacity of the disc near 1~AU 
such that it is unable to match the higher mass accretion rates at larger radii. This can give rise to a limit cycle, in which the 
inner disc alternates between a thermally ionized high state and a low state in which mass is accumulating 
within the dead zone \citep{armitage01}. Bursts are triggered when the growing surface density exceeds the 
threshold for gravitational instability \citep{toomre64}. This gravo-magneto cycle is distinct from the thermal 
instability of dwarf nova accretion discs, but can be characterized locally using an identical {\bc ``S-curve" 
formalism \citep{zhu09a,martin11,martin14}}. A number of one and two-dimensional simulations show that the gravo-magneto 
cycle can lead to global large-amplitude instabilities of protostellar accretion \citep{armitage01,zhu09,zhu10a,martin13,bae13}.

The leading source of uncertainty in gravo-magneto instability models arises from physics that was 
not included in the original dead zone calculations \citep[for reviews, see][]{armitage11,turner14}. 
There is strong evidence for the importance of 
additional non-ideal MHD effects (ambipolar diffusion and the Hall effect) other than Ohmic diffusion \citep{subu13,simon13,lesur14}, 
and substantial indications from local numerical simulations that mass and angular momentum loss 
can occur on AU scales via disc winds \citep{bai13}. {\bc Furthermore, several of the parameters of the instability remain uncertain. The temperature at which thermal ionization takes places has been shown to impact the properties of the bursts, in particular the peak accretion rate reached during the burst \citep{zhu10a}. Additionally, whether or not the dead-zone can support residual levels of turbulent transport remains an open question with simulations suggesting a low level of transport may be possible \citep{fleming03,turner10,okuzumi11}. Such residual transport in the dead-zone has been shown to have an influence on the outburst properties, and hence the evolution of the disc \citep{bae13,bae13b,martin14}. Finally, whether heating from the spiral arms due to the gravitational instability can be considered local and axisymmetric is still an open question \citep{balbus99,gammie01,lodato04} }. These important issues are not, however, our focus here. 
Rather, our goal is to characterize how radial thermal diffusion affects accretion outbursts within a disc 
model subject to a local gravo-magneto instability. Thermal diffusion 
matters because the global evolution of an unstable disc is sensitive not just to the 
details of local instabilities, but also to the radial coupling between annuli. Energy is transported 
radially by advection and by diffusive processes (radiative diffusion, and turbulent transport of heat). 
The role of thermal diffusion in disc outbursts has been studied in the context of dwarf nova outbursts 
and low-mass X-ray binaries \citep{hameury98,lasota01}, but has received little attention in protostellar discs. 
We show that it can substantially affect both the duration of gravo-magneto outbursts, and the 
range of parameter space over which outbursts occur.

The outline of the paper is as follows. In \S2 we describe the general properties of our thermal diffusion 
model, which includes both radiative and turbulent diffusion terms. We argue that the latter is most 
consistently written in terms of the disc's radial entropy (or `potential temperature') gradient. In \S3 we 
describe the one-dimensional disc model, which is used in \S4 to generate a parameter study of 
discs with varying mass infall rates and Prandtl numbers. In \S5 we present analytic estimates 
of the relative importance of radiative and turbulent diffusion terms. \S6 contains our conclusions.

\section{Overview}
In this work we want to investigate the role of {\bc radial} thermal diffusion in the gravo-magneto limit cycle for protoplanetary discs that contain a dead-zone \citep[e.g.][]{gammie96,armitage01,zhu10a,martin13,martin14}, as it has been shown to be important in previous models of accretion disc variability caused by the thermal instability \citep[e.g][]{flp,mineshige86,cannizzo93,hameury98}. Radiative diffusion in the optically thick limit  can be included simply using a `two-stream' approximation for vertical and radial energy transport due to radiation \citep[e.g.][]{flp}. {\bc When the disc is geometrically thin or optically thick vertical radiative diffusion will dominate over radial radiative diffusion}. However, there has been little consensus for how to include thermal diffusion due to turbulence, with different forms existing in the literature. Typically the radial {\bc (vertically averaged)} heat flux due to turbulence is included through a term that is proportional to the temperature gradient \citep[e.g.][]{hameury98}:
\begin{equation}
\tilde{F}_{\rm turb}\propto D^{R}_{\rm th}(R) \frac{\partial T_m}{\partial R}
\end{equation}
where $D^{R}_{\rm th}$ is the diffusion constant  and $T_m$ is the disc's mid-plane temperature. However, as discussed by \citet{balbus00}, { \bc in the context of vertical transport}, turbulent heat flux is not sensitive to the background temperature gradients, but rather the background entropy gradient. This arises since gas parcels moving radially due to the turbulence will move along adiabats over a certain length scale ($\ell$), before exchanging heat with the surroundings. Thus, after the gas parcel has moved a distance $\ell$ the relevant temperature gradient is the one between the gas parcel and the surroundings, which is set by the adiabat along which the gas parcel has moved. Thus, the heat flux will be proportional to the disc's background entropy gradient and the length scale over which the turbulence has moved the gas parcel before it transfers its heat to the surrounding disc. This implies that the term describing the {\bc vertically averaged} heat flux due to turbulence is of the form:
\begin{equation}
\tilde{F}_{\rm turb}\propto D^{R}_{\rm th}(R)\frac{\partial s}{\partial R}\propto\frac{3\nu\Sigma}{Pr}\frac{\partial s}{\partial R}\label{eqn:Fturb}
\end{equation}
where $s$ is the disc's entropy per unit mass,  $\nu$ is the kinematic viscosity, $\Sigma$ is the disc's surface density and the uncertainty in the length scale over which turbulent heat transport takes place has been encapsulated in the non-dimensional Prandtl number ($Pr$), which measures the ratio of angular momentum to heat transport. Equation~\ref{eqn:Fturb} is similar to the form used in planetary atmospheres to capture turbulent transport of heat, where the entropy is often characterised in terms of the `potential temperature' ($\Theta$) \citep[e.g.][]{tritton88}, a form we will adopt in Section~\ref{sec:descp}. {\bc  We note that as is the case for radiative diffusion, the turbulent transport in the vertical direction is likely to exceed the radial transport. The vertical transport may well impact the evolution of disc instabilities \citep{zhunarayan13} but here we focus 
on the radial term. We show that the radial transport of heat by radiative and turbulent diffusion can have a strong impact on the properties of the accretion bursts. } 

\subsection{Basic Thermal Evolution}\label{sec:descp}
As discussed above turbulent diffusion of heat in a protoplanetary disc is likely to be sensitive to the radial entropy gradient in the disc, rather than the radial temperature gradient. Thus, in order to derive an evolution equation for the thermal state of the disc we begin with the entropy equation:{\bc 
\begin{equation}
\rho T\frac{{\rm D}s}{{\rm D}t}=D-\nabla\cdot{\bf F_r}\label{eqn:entropy}
\end{equation}
where $s$ is the specific entropy, $\rho$ is the mass density and $D=9/4\nu\rho\Omega^2$  - with $\Omega$ the Keplerian angular velocity - is the heating rate per unit volume due to viscous dissipation. The final term on the RHS represents heat transport due to radiation, where ${\bf F_r}$ is the radiation flux, which in the diffusion approximation is given by:
\begin{equation}
{\bf F_r}=-\frac{4acT^3}{\kappa\rho}\nabla T
\end{equation}
where $a$ is the radiation constant, $c$ the speed of light and $\kappa$ the opacity.
The entropy is given by:
\begin{equation}
s=C_V\log\left(\Theta\right) + {\rm const.}
\end{equation}
where  $C_V$ is the heat capacity at constant volume and $\Theta$ is a thermodynamic quantity that is conserved during an adiabatic process. By analogy with vertical heat transport in planetary atmospheres, we call $\Theta$ the `potential temperature' \citep[e.g.][]{tritton88} and define $\Theta$ such that it has units of temperature. We assume - as we do for temperature - that it can be defined as a 1D property of the disc, that depends only on radius. The potential temperature is a much easier variable to work with than entropy so we proceed in re-writing Equation~\ref{eqn:entropy} as:
\begin{equation}
\rho \frac{C_V T}{\Theta}\left(\frac{{\rm D}\Theta}{{\rm D}t}\right)=D-\nabla\cdot{\bf F_r}
\end{equation}

\subsection{Including turbulent heat transport}
We may introduce an phenomenological term for the turbulent diffusion of potential temperature as:
\begin{equation}
\rho \frac{C_V T}{\Theta}\left(\frac{{\rm D}\Theta}{{\rm D}t}+\frac{\nabla\cdot{\bf F}_\Theta}{\rho}\right)=D-\nabla\cdot{\bf F_r}\label{eqn:PT}
\end{equation}
where ${\bf F}_\Theta$ is the diffusive flux of potential temperature due to turbulence. The entropy equation including a turbulent heat flux is then given by:
\begin{equation}
 \rho T\frac{{\rm D}s}{{\rm D}t}=D-\nabla\cdot{\bf F_r}-\frac{C_VT}{\Theta}\nabla\cdot{\bf F_\Theta}\label{eqn:entropy2}
\end{equation}
Under the assumption of an ideal gas, and neglecting compressional heating due to $P{\rm d}V$ work which is negligible for the velocities in protoplanetary discs, Equation~\ref{eqn:entropy2} can be written in an explicit form for the evolution of the temperature as:
\begin{equation}
C_p\rho\frac{DT}{Dt}-D+\nabla\cdot{\bf F_r}=-\frac{C_VT}{\Theta}\nabla\cdot{\bf F_\Theta} \label{eqn:temp2}
\end{equation}
where $C_p$ is the heat capacity at constant pressure. To obtain the 1D disc equation we vertically integrate Equation~\ref{eqn:temp2} following \citet{flp,cannizzo93}, where the terms of the LHS are vertically averaged in the standard way, with
\begin{equation}
\int_{-\infty}^{\infty}\!\!\!\!\!{\rm d}z\,C_p\rho\frac{DT}{Dt}\approx C_p\Sigma\frac{DT_m}{Dt}
\end{equation}
and
\begin{equation}
\int_{-\infty}^{\infty}\!\!\!\!\!{\rm d}z\,\nabla\cdot{\bf F_r}\approx\frac{16}{3\tau}\sigma_bT_m^4+\frac{2H}{R}\frac{\partial}{\partial R}\left(R\frac{4acT_m^3}{\kappa \rho_m}\frac{\partial T_m}{\partial R}\right)
\end{equation}
where $\rho_m$ is the mid-plane density of the disc, $\sigma_b$ is the Stefan-Boltzmann constant, $H$ is the disc's scale height and $\tau=\Sigma\kappa/2$ is the vertical optical depth. Finally, we assume there is no vertical turbulent energy flux through the surface of the disc and vertically integrate the radial flux term for the potential temperature following \citet{mineshige86,cannizzo93,hameury98}\footnote{Note, these authors use a vertically averaged flux term that depends on temperature, rather than potential temperature; however, the procedure is identical.}, such that:
\begin{equation}
\int_{-\infty}^{\infty}\!\!\!\!\!{\rm d}z\,\frac{C_VT}{\Theta}\nabla\cdot{\bf F_\Theta}\approx\frac{C_V T_m}{\Theta_mR}\frac{\partial}{\partial R}\left(R\tilde{F}_\Theta\right) 
\end{equation}
where $\Theta_m$ is the mid-plane potential temperature, and $\tilde{F}_\Theta$ is the vertically averaged turbulent heat flux. Thus the vertically averaged equation for the evolution of the mid-plane temperature, including turbulent heat flux becomes:
\begin{eqnarray}
\frac{\partial T_m}{\partial t}&=&-v_R\frac{\partial T_m}{\partial R}+\frac{1}{\Sigma}\left[\frac{\Gamma}{C_p}-\frac{T_m}{\gamma\Theta_m R}\frac{\partial}{\partial R}\left(R\tilde{F}_\Theta\right)\right]\nonumber\\&&-\left(\frac{2H}{\Sigma R {\bc C_p}}\right)\frac{\partial}{\partial R}\left(R\frac{4acT_m^3}{\kappa \rho_m}\frac{\partial T_m}{\partial R}\right)\label{eqn:temp}
\end{eqnarray}
where $\Gamma/2$ is the net heating rate per unit area, $\gamma=C_p/C_V$ is the ratio of heat capacities.  We have manipulated  the energy equation into this form as our 1D viscous evolution code already contains routines to integrate the temperature equation \citep{owen14}  as well as being similar to those used in other studies \citep{cannizzo93,armitage01,zhu10a,zhu10b}. This allows for consistency within our numerical method as well as allowing us to easily switch on and off the additional diffusive transport terms of interest. Without radiative and turbulent diffusion Equation~\ref{eqn:temp} is identical to that used by \citep{armitage01} to study bursts in the dead-zone accretion scenario of protoplanetary discs and similar to those used by \citet{zhu10a,zhu10b,martin13,martin14}. Ultimately, the turbulent heat transport term will only be important when the disc is out of local thermal equilibrium, i.e. when instabilities arise from the activation or  quenching of material in the dead zone which results in accretion bursts.  

}
\subsubsection{Choice of `Potential Temperature'}
Given we want our choice of potential temperature to be invariant under an adiabatic processes that moves a fluid element in a disc radially we start from the standard adiabatic relation $T\rho^{1-\gamma}={\rm const}$. Thus, replacing $\rho_m\propto\Sigma/H$ and remembering $H\propto T_m^{1/2}/\Omega$ where $\Omega$ is the angular velocity of the gas  (taken to be Keplerian), we find an appropriate form for {\bc the mid-plane} potential temperature for an accretion disc is:
\begin{equation}
\Theta_m\propto T_m\left(\Sigma\Omega\right)^{\frac{2(1-\gamma)}{1+\gamma}}
\end{equation}
where we note the form of the {\bc vertically averaged} turbulent heat transport term in Equation~\ref{eqn:temp} is independent of the chosen scaling for the potential temperature. To prevent overflow/underflow errors in our numerical model described in Section~\ref{sec:num} we choose to normalise the surface density and angular velocity to the Minimum Mass Solar Nebula values at 1~AU.

Furthermore, by dimensional arguments coupled with comparisons with the angular momentum flux due to the turbulence we define the {\bc vertically averaged} potential temperature flux as (c.f. \citealt{mineshige86,cannizzo93,hameury98}):
\begin{equation}
\tilde{F}_\Theta=-\frac{3\nu\Sigma}{Pr}\frac{\partial \Theta_m}{\partial R}
\end{equation}
Where the most naive expectation would be $Pr=1$. 

\section{one-dimensional disc model}
In order to investigate the effect of turbulent heat diffusion on accretion burst that can occur in protoplanetary discs with dead zones we numerical integrate the disc equations using a simple one-layer approach. Thus, following \citet{armitage01} we integrate the diffusion equation for the surface density for a disc around a solar mass star:
\begin{equation}
\frac{\partial\Sigma}{\partial t}=\frac{3}{R}\left\{R^{1/2}\frac{\partial}{\partial R}\left[\left(\nu_{\rm MRI}\Sigma_{\rm ac}+\nu_{\rm SG}\Sigma\right)R^{1/2}\right]\right\}+\dot{\Sigma}_{\rm in}\label{eqn:gas}
\end{equation}
where $\nu_i$ is the kinematic viscosity due to the MRI and/or self-gravity which we detail in Section~\ref{sec:viscosity} and $\dot{\Sigma}_{\rm in}$ is the rate of material falling onto the disc. $\Sigma_{\rm ac}$ is the surface density that is `MRI active' and is given by:
\begin{equation}
\Sigma_{\rm ac}=
\begin{cases} \Sigma & \textrm{if\,\,} T_m > T_{\rm crit} \textrm{\,\,or\,\,} \Sigma < 2\Sigma_{\rm layer} \\
2\Sigma_{\rm layer} & \textrm{otherwise}
\end{cases}
\end{equation}
For all our models we choose $T_{\rm crit}=800\,$K and $\Sigma_{\rm layer}=10^{2}\,$g~cm$^{-2}$.

The temperature evolution of the disc is given by Equation~\ref{eqn:temp}
where the net heating rate ($\Gamma$) is given by:
\begin{equation}
\Gamma=\frac{9}{4}\nu\Sigma_{\rm ac}\Omega^2-\frac{16}{3\tau}\sigma_b T_m^{4}
\end{equation}
For our model we adopt the \citet{belllin97} opacities. Finally, following \citet{armitage01} when a region of the disc makes the transition to the layered state, we do not attempt to treat the vertical structure but simply replace the expression for the optical depth by $\tau=\Sigma_{\rm layer}\kappa$. 

In all calculations we assume an idea equation of state with $\gamma=7/5$ and a mean-molecular weight ($\mu$) of 2.35 as the majority of disc material we are interested in remains molecular and neutral. 

\subsection{Viscosity and Self-Gravity}\label{sec:viscosity}
We treat the viscosity using an `alpha' model such that:
\begin{equation}
\nu_i=\alpha_i c_s H
\end{equation}
where $c_s$ is the isothermal sound speed. For MRI turbulence we set $\alpha_{MRI}=0.01$ and for angular momentum transport due to self-gravity {\bc we follow \citet{lin_pringle87,armitage01}} and set $\alpha_{\rm SG}$ to
\begin{equation}
\alpha_{\rm SG}=\begin{cases} 0.01\left(\frac{Q^2_{\rm crit}}{Q^{\bc 2}}-1\right) & \mbox{if $Q\le Q_{\rm crit}$} \\
0 & \mbox{if $Q> Q_{\rm crit}$} \end{cases}\label{eqn:alpha_SG}
\end{equation}
{\bc where $Q_{\rm crit}=2$} and $Q$ is the Toomre parameter \citep{toomre64} given by:
\begin{equation}
Q=\frac{c_s\Omega}{\pi G \Sigma}
\end{equation}
The form of $\alpha_{SG}$ is not important and is merely designed to keep $Q\sim Q_{\rm crit}$ \citep[e.g.][]{lin_pringle87,lin_pringle90,zhu10a} as suggested by simulations of self-gravitating discs \citep[e.g.][]{lodato04,lodato05}.

\subsection{Numerical Method}\label{sec:num}
Our numerical code is based on the method detailed in \citet{owen14}; Equations~\ref{eqn:temp} \& \ref{eqn:gas} are integrated explicitly using a scheme that is second order in space and first order in time, where advective fluxes are reconstructed using a second order method and a Van-Leer limiter. We use a non-uniform radial grid {(\bc uniformly spaced in $R^{7/2}$)} with 200 cells with an inner boundary at $0.047$~AU  and an outer boundary at $66.7$~AU. {\bc The chosen grid has high enough resolution so that the results are converged.}.  At the inner boundary we apply a zero shear boundary condition for the surface density and a reflection boundary condition for the temperature. At the outer boundary we apply a zero mass-flux boundary condition for the surface density and a reflection boundary condition for the temperature. We initialise the simulations with the gas temperature at 10~K, and the gas surface density set to the numerical floor value $\Sigma=10^{-20}$~g~cm$^{-3}$. Additionally the gas temperature is not allowed to fall below 10~K at any time during the integration. Material then falls onto the disc at a steady value $\dot{M}_{\rm in}$, where $\dot{\Sigma}_{\rm in}$ is taken to be a Gaussian profile of width 1~AU and centred at 10~AU, (outside the location of any dead-zone). We then evolve the system to 5~Myrs, over which time either a steady disc profile with an accretion rate profile $\dot{M}(R)=\dot{M}_{\rm in}$ or a steadily repeating limit cycle is achieved, typically this takes 0.5-2~Myrs to reach. Mean burst properties are determined by averaging the simulation results from 2-5~Myrs. 

\section{Results}
\begin{figure}
\centering
\includegraphics[width=\columnwidth]{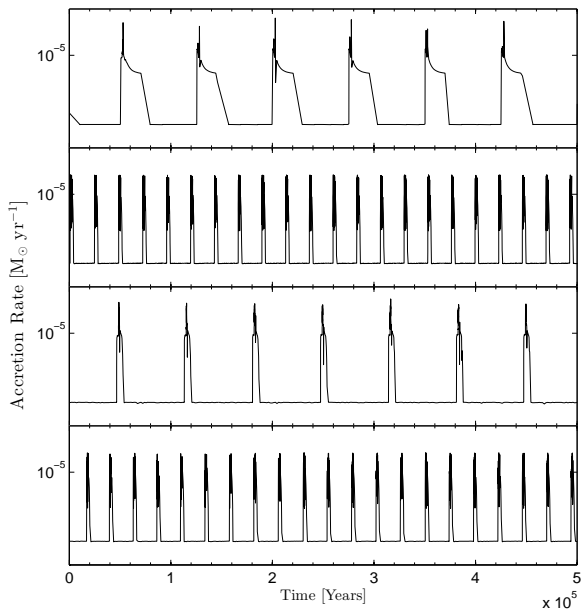}
\caption{Time evolution of the mass-accretion rate at the inner boundary with an infall accretion rate of $1.4\times10^{-6}$~M$_\odot$~yr$^{-1}$, the origin of the x-axis has been reset to zero after a steady limit cycle has been reached. The top panel shows a simulation with no radiative diffusion and no turbulent diffusion. The second panel shows a simulation with no radiative diffusion and turbulent diffusion with $Pr=1$. The third panel shows a simulation with radiative diffusion and no turbulent diffusion and the bottom panel show a simulation with both radiative diffusion and turbulent diffusion with $Pr=1$.}\label{fig:demo}
\end{figure}

We perform a series of simulations where we vary the input infall rate, Prandtl number and whether or not radial radiative diffusion is included. In Figure~\ref{fig:demo}, we show the temporal evolution of the mass-accretion rate onto the star ($\dot{M}_*$) for an infall rate of $\dot{M}=1.4\times10^{-6}$~M$_\odot$~yr$^{-1}$, where we switch on and off the different thermal diffusion effects. We switch off radiative diffusion entirely for simulations with no radiative diffusion, for cases with no turbulent diffusion we just set the Prandtl number to $10^6$. 

The top panel shows a simulation with no radial diffusion. The second panel shows a simulation with turbulent diffusion ($Pr=1$) and no radiative diffusion. The third panel shows a simulation with radiative diffusion but no turbulent diffusion and the final panel shows a simulation with both radiative and turbulent diffusion ($Pr=1$).  Figure~\ref{fig:demo} clearly shows that thermal diffusion has a strong impact on the properties of the gravo-magnetic limit cycle, where thermal diffusion shortens the burst duration significantly from $\sim2\times10^4$ years with no diffusion to $\sim10^{3}$ years in the case of turbulent diffusion. Figure~\ref{fig:demo} also indicates that while radiative diffusion alone decreases the burst duration in does not increase the frequency of bursts significantly, whereas the addition of turbulent diffusion also increases the burst frequency. Finally, we note that with strong turbulent diffusion ($Pr=1$) it dominates over radiative diffusion, and the simulation with both radiative and turbulent diffusion is almost identical to the the simulation with just turbulent diffusion. 

\subsection{Parameter Study}

Since the mass-accretion rate in protoplanetary discs varies by many orders of magnitude during the evolution of a protoplanetary disc \citep[e.g.][]{hartmann98,ercolano14} and the Prandtl number is an unknown parameter, in the next set of simulations we vary both the infall rate onto the disc from $10^{-4}-10^{-9}$~M$_\odot$~yr$^{-1}$ and the Prandtl number from $10^{4}-0.1$ in a series of simulations where we include both radiative diffusion and turbulent diffusion. Most simulations show the time evolution shown in Figure~\ref{fig:demo}, where the disc quiescently accretes through the active surface layers above the dead-zone at $\dot{M}_*\sim10^{-8}$~\msunyr. The dead-zone is then heated above the activation temperature and the mass-accretion rates increases dramatically, with the star accreting at $\gtrsim 10^{-4}$~\msunyr for a short period before the disc returns to quiescence.

In the majority of our calculations the disc temperature remains above the value expected for a passively heated disc \citep[e.g.][]{kenyon87,CG97,dalessio01} in the  region of interest; justifying our use a purely viscously heated disc. However, we note for the lowest infall rates $\lesssim 10^{-8}$ \msunyr the disc's temperature drops below the \citet{CG97} passively heated temperature profile outside $\sim 2$~AU. Therefore, in-order to model the full evolution of a disc (something we do not attempt here) then heating of the disc due to stellar irradiation will need to be taken into account. Since, the aim of this work is purely to analyse the effect of thermal diffusion on gravo-magnetic limit cycle, neglecting stellar irradiation (for the small range of parameters where it maybe important) does not affect our results or the inferences we draw from them. 

\begin{figure}
\centering
\includegraphics[width=\columnwidth]{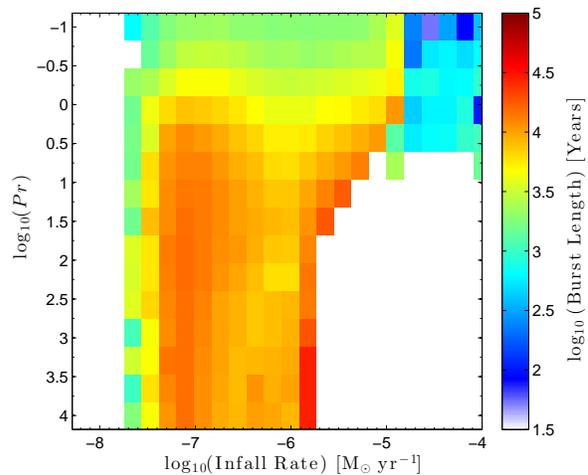}
\caption{The burst length shown as a function of infall rate and Prandtl number for simulations with no radiative diffusion. White regions indicate those regions which do not show a limit cycle, but rather a constant accretion rate through the disc which matches the infall rate, we note the noise at low accretion rates is due to very few burst being present during the simulation.}\label{fig:b_length1}
\end{figure}

\begin{figure}
\centering
\includegraphics[width=\columnwidth]{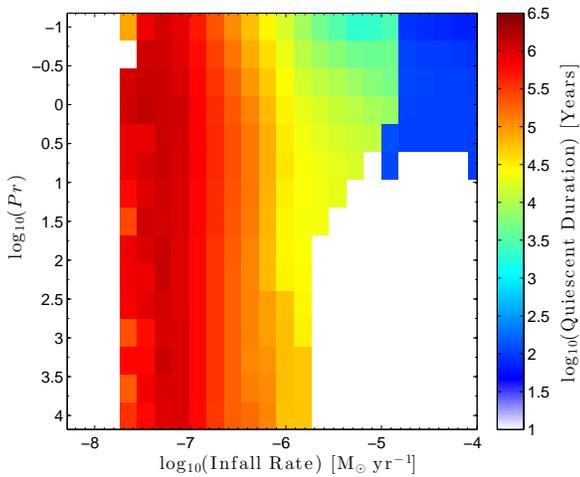}
\caption{Length of quiescent periods in between bursts shown as a function of infall rate and Prandtl number. White regions indicate those regions which do not show a limit cycle, but rather a constant accretion rate through the disc which matches the infall rate.}\label{fig:b_freq1}
\end{figure}

\begin{figure}
\centering
\includegraphics[width=\columnwidth]{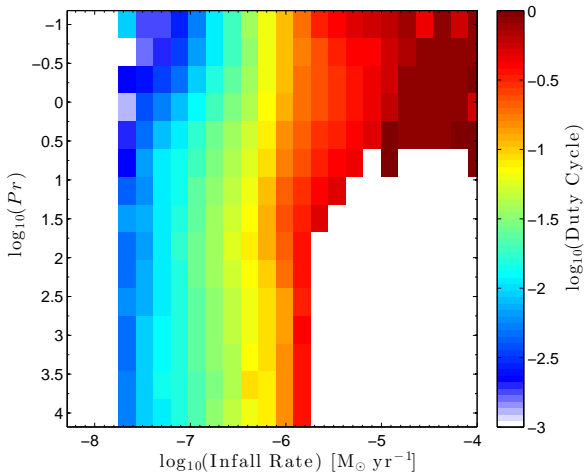}
\caption{The duty cycle of the bursts shown as a function of infall rate and Prandtl number. White regions indicate those regions which do not show a limit cycle, but rather a constant accretion rate through the disc which matches the infall rate.}\label{fig:duty_1}
\end{figure}

 We use the simulations to measure several parameters of interest including the average burst duration, the average time between bursts and the duty cycle (the percentage of time the disc spends in a burst). In Figures~\ref{fig:b_length1}, \ref{fig:b_freq1} \& \ref{fig:duty_1} we show the burst length, time between bursts and duty-cycle for the set of simulations without radiative diffusion. 

These results show that for Prandtl numbers less than $\sim 1000$ the limit-cycle properties are significantly altered.  The general trend discussed previously that turbulent diffusion leads to significantly shorter bursts that occur more frequently is seen for the full range of parameters. With Prandtl numbers of order unity the burst duration is approximately an order of magnitude shorter - $\sim 1000$ years - compared to simulations with low levels of turbulent diffusion. Additionally the range of accretion rates that are unstable to the limit cycle increases at higher accretion rates, where without turbulent diffusion discs with infall rates $\gtrsim 2\times10^{-6}$~\msunyr are stable to bursts; however, with Prandtl numbers of order unity this appears to extend for the full simulated range upto infall rates of $10^{-4}$~\msunyr. While we see the individual burst properties appear to be significantly altered, the duty-cycle is not as greatly changed where in general even with the lowest Prandtl numbers it within a factor of two of the value without turbulent diffusion. 

\begin{figure}
\centering
\includegraphics[width=\columnwidth]{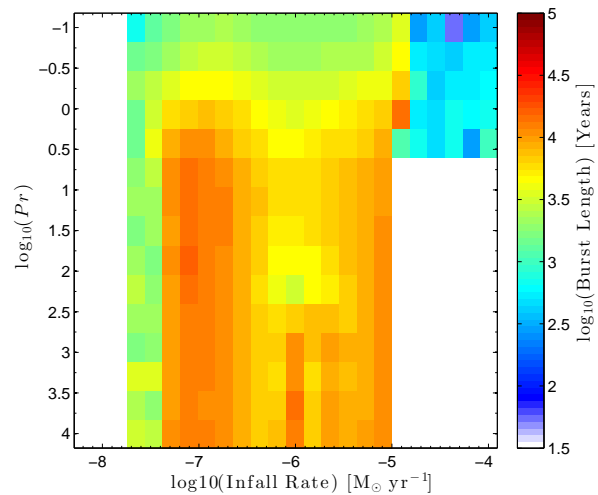}
\caption{Same as Figure~\ref{fig:b_length1}, but for simulations that include radiative diffusion}.\label{fig:b_length2}
\end{figure}

\begin{figure}
\centering
\includegraphics[width=\columnwidth]{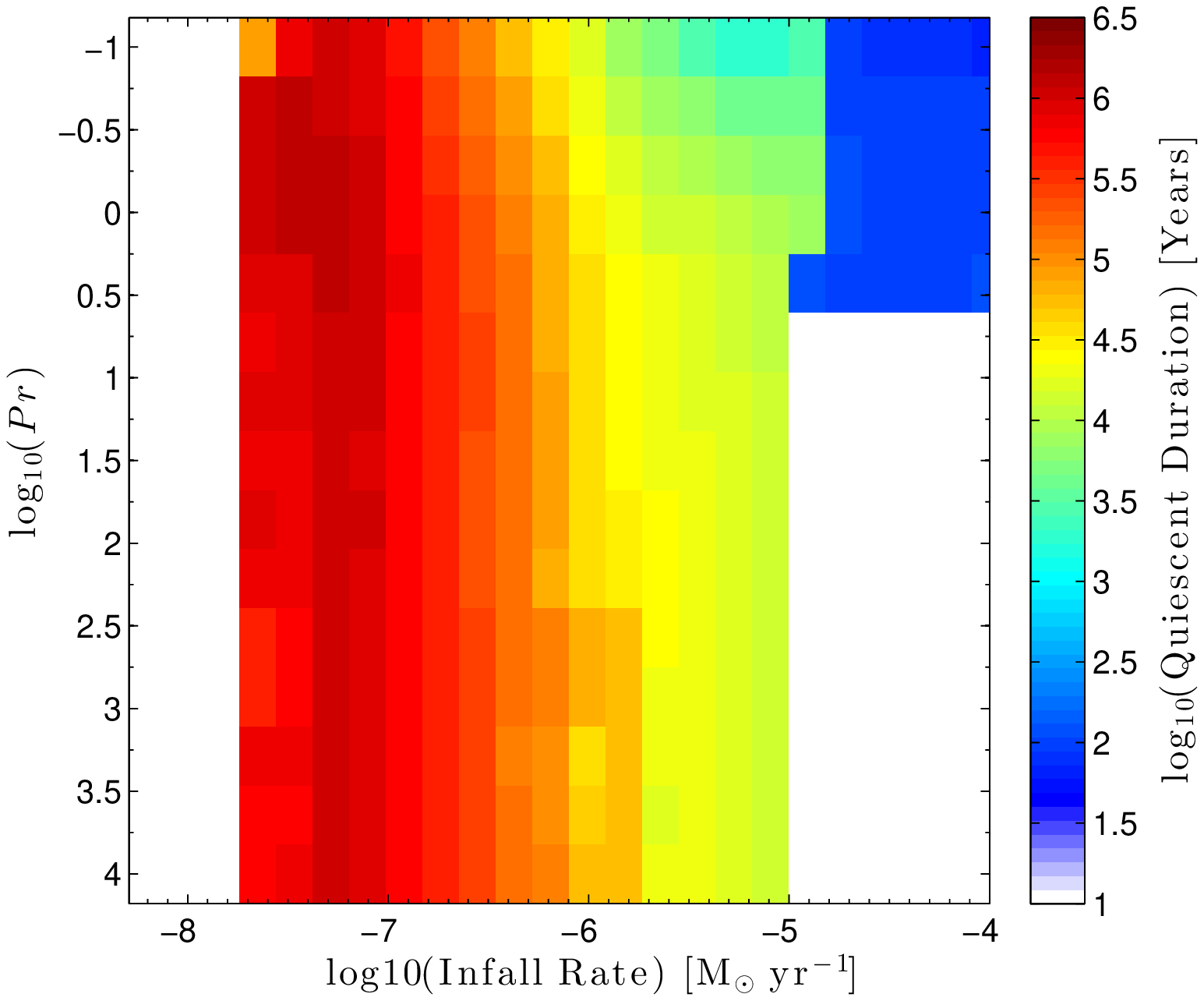}
\caption{Same as Figure~\ref{fig:b_freq1}, but for simulations that include radiative diffusion.}\label{fig:b_freq2}
\end{figure}

\begin{figure}
\centering
\includegraphics[width=\columnwidth]{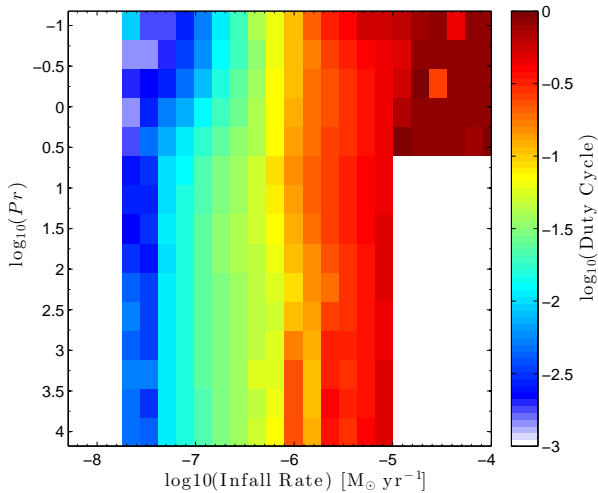}
\caption{Same as Figure~\ref{fig:duty_1}, but for simulations that include radiative diffusion.}\label{fig:duty_2}
\end{figure}

In Figures~\ref{fig:b_freq2}, \ref{fig:b_length2} \& \ref{fig:duty_2} we show the same information on burst length, frequency and duty-cycle but for the set of simulations that additionally include radiative diffusion. Again these simulations show the same general features as expected from the results discussed above. Thermal diffusion leads to shorter burst durations. Additionally the range of infall rates over which accretion bursts occur is extended to higher accretion rates, with burst presents at accretion rates an order of magnitude higher at $\sim10^{-5}$~\msunyr at high Prandtl numbers.  However, it is clear that while radiative diffusion is important if turbulent diffusion is unimportant for Prandtl numbers smaller than $\sim 25$ the similarities between the simulations with and without radiative diffusion indicate that turbulent diffusion dominates the transport of heat over radiative diffusion. 

\begin{figure}
\centering
\includegraphics[width=\columnwidth]{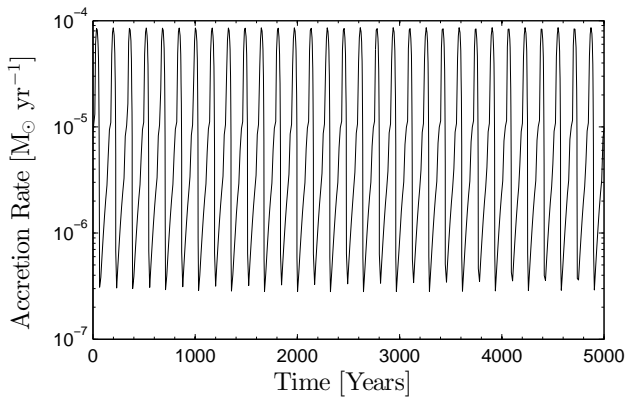}
\caption{Time evolution of a the mass accretion rate onto the star during a simulation with an infall rate of $1.86\times10^{-5}$~\msunyr and $Pr=0.518$. The time has been reset to zero after a steady cycle was achieved. }\label{fig:fast}
\end{figure}

Finally, we note at small Prandtl numbers $\lesssim 10$ and high infall rates $\gtrsim 10^{-5}$~\msunyr the properties of the burst are different from the standard burst profile. The mass-accretion rate as a function time is shown for one such case with an infall rate of $1.86\times10^{-5}$~\msunyr and $Pr=0.518$ in Figure~\ref{fig:fast}. Where the disc does not return to a quiesent level of accretion at a rate of $10^{-8}$~\msunyr, but rather rapidly oscillates between an accretion rate of $\sim 5\times10^{-7}$ and $10^{-4}$~\msunyr. 

\section{Discussion}
We have shown that radial diffusion of heat by turbulence (or radiation) has a significant impact on the length and frequency of accretion burst due to the gravo-magnetic limit cycle arising from dead-zones in protoplanetary discs. In particular, we find that turbulent diffusion with Prandtl numbers $\lesssim 1000$ significantly shortens the burst duration and increases the burst frequency. With Prandtl numbers of order unity the burst length can be upto an order of magnitude shorter than previously reported values \citep[e.g.][]{armitage01,zhu10a}. 

{\bc Our simulations occasionally show the onset of thermal instability during the `high-state' (the variability in the accretion rate during the burst shown in Figure~\ref{fig:demo} is indicative of this), as has been seen in other simulations \citep{zhu09,zhu10a}. Radial heat diffusion does modify the thermal instability, an effect that has been studied in detail previously \citep{flp,cannizzo93,ludwig98,hameury98}.
}
\subsection{Turbulent versus radiative diffusion}
\begin{figure}
\centering
\includegraphics[width=\columnwidth]{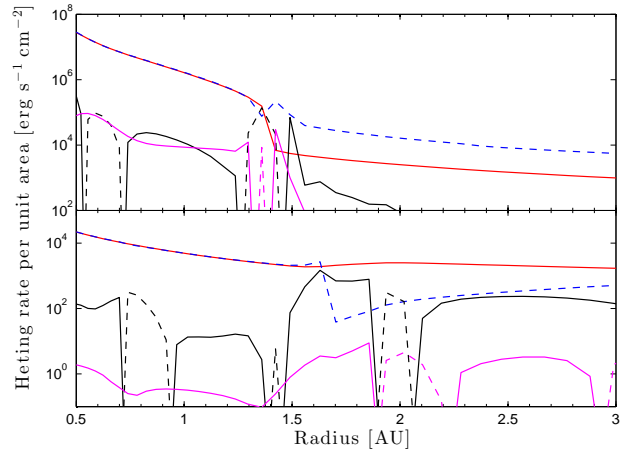}
\caption{Heating/cooling rates per unit area (vertically integrated heating rates) as a function of disc radii, solid lines indicate net heating and dashed-lines indicate net cooling. The red line shows viscous heating, blue shows vertical cooling, black shows turbulent diffusion and magenta shows radiative diffusion. The top panel is for a disc undergoing a burst and the bottom panel shows the disc during quiescence.}\label{fig:rates}
\end{figure}

One of the clearest results from the simulations is that at $Pr\lesssim 25$ turbulent diffusion dominates over radiative diffusion in setting the burst properties. This in contrast to previous results on the thermal instability which suggested the terms had a similar role for $Pr=1$ \citep{cannizzo93}. {\bc Thus, radiative and turbulent diffusion play different roles when there are large temperature or viscosity gradients.}  Noting, that in thermal equilibrium one can calculate the `Prandtl number' for radiative diffusion ($Pr^{\rm rd}$)  \citep{paardekooper11} -- finding ${\bc Pr^{\rm rd}}\sim 1$ in the optically thick limit -- it may seem surprising that  in the simulations turbulent diffusion dominates over radiative diffusion. In order to investigate this further we plot the various heating rates during quiescence (bottom) and during the burst (top) in Figure~\ref{fig:rates} for the simulation shown in Figure~\ref{fig:demo} (bottom panel) with an infall rate of $1.4\times10^{-6}$~\msunyr and Prandtl number of 1 including radiative diffusion. We note in all cases heating/cooling due to advection is negligible.

This indicates that turbulent diffusion does indeed dominate the heating/cooling in regions where the disc is out of local thermodynamic equilibrium. The main difference between the gravo-magnetic burst discussed here and the thermal instability is that in this situation temperature is rather continuous through the front, but we have large changes in the viscosity. Whereas in the thermal stability there are larger changes in temperatures across the  instability front which also gives rise to large changes in the viscosity. Since the thermal `diffusivity' constant in the turbulent transport case is sensitive the viscosity, whereas the thermal `diffusivity' constant in the radiative diffusion case is sensitive to temperature, then for large changes in viscosity across the dead-zone and a comparatively continuous temperature distribution the heating rate for turbulent diffusion is going to dominate over the heating rate for radiative diffusion. Whereas, in the case of the thermal instability they are likely to be comparable across the heating front as suggested by previous simulations. Finally, we note that since heat transport due to thermal diffusion is only likely to dominate out of thermal equilibrium and the high power of $T$ in the radiative diffusion `diffusivity' constant means that an analysis assuming thermodynamic equilibrium can only crudely capture the results; something we discuss further in Section~\ref{sec:PR_est}. 

{\bc Finally, we note that the 2D simulations of \citet{zhu09} indicated that when the thermal instability was triggered during an outburst it could lead to vigorous convection. This process could certainly be affected by the radial transport seen here and further modelling is warranted. }

\subsection{Role of thermal diffusion}
Diffusion of heat by turbulence or radiation  plays a major role when the disc is out of local thermal equilibrium (i.e. when viscous heating is not balanced by vertical cooling). This can occur in the dead zone; prior to the burst, viscous heating is stronger than radiative cooling  (which results in a temperature increase - triggering the burst) and during the burst, vertical radiative cooling is stronger than viscous heating (which results in a temperature decrease - shutting off the burst). 

\begin{figure}
\centering
\includegraphics[width=\columnwidth]{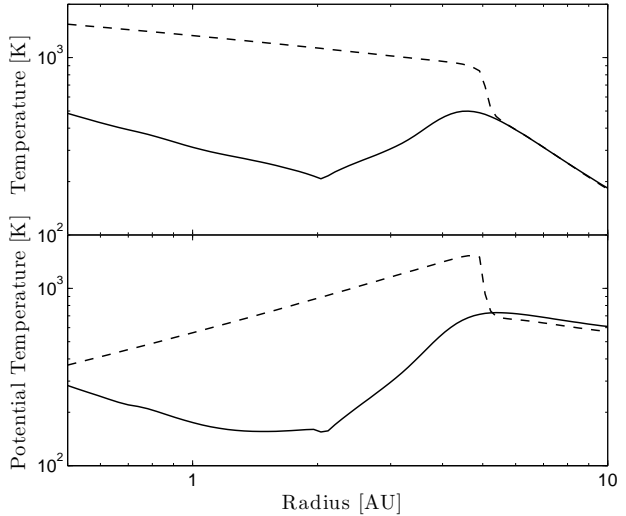}
\caption{Temperature (top) and potential temperature (bottom) panel structure of the dead-zone region shown during quiescence (solid) and during the burst (dashed). We have scaled the potential temperature to surface densities and angular velocities of the Minimum Mass Solar Nebula so the units are sensible.}\label{fig:temp_dead}
\end{figure}

In Figure~\ref{fig:temp_dead}, we show the temperature (top) and potential temperature (bottom) structure in the dead-zone region during quiescence (solid) and during the burst (dashed). This indicates that during quiescence thermal diffusion (turbulent and radiative) will heat-up the dead-zone region from both sides triggering the burst faster. During the burst, the dead zone region is hotter than the surrounding disc - at the outer edge for temperature (thus important for radiative diffusion)  and both sides for potential temperature (thus important for turbulent diffusion) - thus thermal diffusion will cool the dead-zone region shutting off the burst quicker. 

We can gain some insight by calculating the relevant thermal time-scales for different processes. Taking the viscous time-scale to be $t_\nu=R^2/3\nu$, then the heating time-scale ($t_{\rm heat}$) due to viscous processes is approximately:
\begin{equation}
t_{\rm heat}\approx\left(\frac{H}{R}\right)^2t_\nu
\end{equation}
whereas the cooling time ($t_{\rm cool}$) can be expressed as:
\begin{equation}
t_{\rm cool}=\left(1+\frac{\delta T}{T_{\rm eq}}\right)^{-4}t_{\rm heat}
\end{equation}
where $\delta T=T_m-T_{\rm eq}$ and $T_{\rm eq}$ is the temperature of the disc in local thermal equilibrium. 

Whereas the time-scale for heating/cooling due to diffusion ($t_{\rm diff}$, turbulent or radiative) is given by:
\begin{equation}
t_{\rm diff}=Pr\left(\frac{l}{R}\right)^2 t_\nu
\end{equation}
where $l$ is the length scale over which the diffusive fluxes vary. Figure~\ref{fig:temp_dead} indicates at the dead-zone boundaries $\ell \ll R$ for the length-scale over which the temperature/potential temperature varies, indicating $t_{\rm diff}\lesssim t_{\rm heat}\mbox{ or }t_{\rm cool}$ and that heating/cooling due to diffusive can become dominant. Furthermore, as discussed above in the case of turbulent diffusion there can be a large viscosity jump across the dead-zone boundary, in this case $l < H$, explaining why turbulent diffusion can be a dominant heat transport processes for low Prandtl numbers $Pr\lesssim 25$.  

Therefore, the results of our simulations can easily be interpreted. During quiescence the dead-zone is cooler than the surrounding disc and out of thermal equilibrium; thermal diffusion heats up the dead-zone region by transporting heat into it, thus, triggering the burst quicker than without thermal diffusion. During the burst the dead-zone is hotter than the surrounding disc (particular when considered in terms of potential temperature); thermal diffusion removes heat from the dead-zone shutting off the burst quicker. As such the burst length and quiescent length is both shortened, resulting in a more rapidly varying limit cycle.  

{\bc The effect of radial transport through the disc on the burst cycle is shown in Figure~\ref{fig:temp_cycle} where we plot the temperature profile through the disc at four stages of the burst cycle: just before ($\sim 100$~years) the burst is triggered (first panel); during the burst (second panel); just before ($\sim 100$~years) the burst ends  (third panel) and just after the disc has returned to quiescence (fourth panel). The solid lines shows a simulation with a Prandtl number of $10^{4}$ and the dashed line shows a simulation with a Prandtl number of $0.1$, both simulations have an infall rate of  $1.4\times10^{-6}$~\msunyr.

\begin{figure*}
\centering
\includegraphics[width=\textwidth]{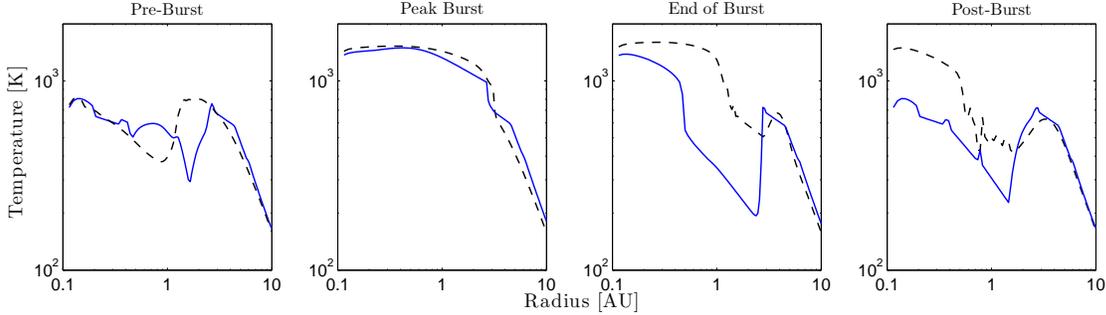}
\caption{The temperature profile for simulations with $Pr=10^{4}$ (solid) and $Pr=0.1$ (dashed) plotted at four stage during an accretion burst: just before the burst is triggered (left panel); during the burst (second panel); just before the burst ends (third panel) and finally when the disc returns to quiescence (fourth panel). Both simulations have an infall rate of $1.4\times10^{-6}$~\msunyr. }\label{fig:temp_cycle}
\end{figure*}

We see that during most stages of the burst's evolution the dead-zone region ($\sim1-5$~AU) is hotter in the simulations with strong radial transport of heat compared to the simulation without strong radial transport of heat. Only during the burst are the temperature profiles very similar, due to the fact that the disc is fully active throughout the dead-zone and close to radiative equilibrium (negating the influence of radial transport). The fact that the dead-zone region is hotter (due to radial transport of heat into the dead-zone from the surroundings) means that it takes a shorter time and less viscous heating (from less disc material) to heat the dead-zone above the activation temperature (800~K) and trigger the accretion burst. The requirement of less heating from viscous dissipation due to heating from radial transport results in the faster burst cycle seen in the simulations with lower Prandtl numbers.
}
\subsection{Estimate of the Prandtl Number for Turbulent diffusion}\label{sec:PR_est}

Since we have indicated that Prandtl numbers $\gtrsim 1000$ are important for affecting the properties of the accretion bursts it is useful to estimate what kind of Prandtl numbers one might expect in protoplanetary discs. In order to approach this we need to assume local radiative equilibrium (i.e. when thermal diffusion is unimportant for disc evolution); however, it should indicate whether the Prandtl number is of the order such that it is important or not. Thus, formally defining the Prandtl number as the ratio angular momentum transport to heat diffusion in the radial direction ($D^R_{\rm th}$) as:
\begin{equation}
Pr=\frac{\alpha c_s^2\Omega^{-1}}{D^R_{\rm th}}
\end{equation}
We consider a mixing length (MLT) approach where heat is transported by eddies of size $\ell$, velocity $v_\ell$ and over-turn time $t_\ell$. If the time-scale for a gas parcel in the eddy to come into thermal equilibrium (due to energy transport by radiation, $t_r$) is much longer than the eddy turn-over time at that scale then, such an eddy is able diffuse heat over the scale $\ell$. Alternatively if $t_r$ is much shorter than the eddy overturn time then the eddy will remain in thermodynamic equilibrium with surroundings and is unable to diffuse heat. Thus in considering the range of possible Prandtl numbers one must evaluate $t_r(\ell)$ and compare it to $t_\ell$. For Kolmogorov turbulence $t_\ell=t_{\rm eddy}(\ell/\ell_{\rm eddy})^{2/3}$ where $t_{\rm eddy}$ and $\ell_{\rm eddy}$ are the time and length at the inertial scale of the turbulence respectively. 

One can estimate $t_r$ following a mixing length style argument (modified from the MLT derivation for convection in stars e.g. - \citealt{kw_book}). Consider a fluid parcel of size $\sim \delta zR\delta\phi\delta R$ that has been displaced from its original location by an eddy of scale $\ell\sim \delta R/2$. Then the radiative flux from the fluid parcel - assuming the radiation field is optically thick - is given by:
\begin{equation}
F_b=\left|\frac{c}{3\kappa_b\rho_b}\nabla\left(aT_b^4\right)\right|
\end{equation}
where the subscript $b$ indicates the properties of the fluid parcel, rather than the surroundings. Assuming the temperature difference between the fluid parcel and surroundings is $\Delta T$ and taking $\nabla T_b\sim 2\Delta T/\ell$ then we find:
\begin{equation}
F_b=\frac{8acT_b^3}{3\kappa_b\rho_b}\frac{\Delta T}{\ell}
\end{equation}
Therefore, the thermal energy transferred between the fluid parcel and surroundings in $t_r$ is:
\begin{equation}
E_{\rm tran}=\Delta T\frac{8acT_b^3}{3\kappa_b\rho_b}\frac{R\delta\phi\delta z}{\ell} t_r
\end{equation}
Equating this to the thermal energy difference between the fluid parcel and surroundings
($C_p\rho_b\delta zR\delta\phi\delta R\Delta T$), one can obtain an expression for $t_r$ in terms of the eddy size $\ell$:
\begin{equation}
t_r(\ell)=A\frac{C_p\rho_b^2\kappa_b\ell^2}{12acT_b^3}\label{eqn:t_r_ell}
\end{equation}
where A is an order unity geometry factor describing the shape of the `fluid' parcel and we have adopted the form factor describing the shape of the fluid parcel from \citet{kw_book} to evaluate Equation~\ref{eqn:t_r_ell}. 

Since $t_r\propto \ell^2$ and $t_\ell \propto \ell^{2/3}$ then the ratio of the ratio of the eddy overturn time to the time-scale on which the eddy reaches thermodynamic equilibrium scales as $t_\ell/t_r\propto \ell^{-4/3}$. This scaling means that as one moves to smaller scales in the turbulent cascade the eddies reach radiative equilibrium with their surroundings faster. 

This result implies that whatever $t_r$, heat transport is most likely to be dominated at the inertial scale, {\bc since the largest eddies maintain their entropy for longer - by reaching thermodynamic equilibrium with the background slower - and move larger distances}. Thus, the ratio of most interest is $\tau_r/t_{\rm eddy}$, where $\tau_r=t_r(\ell_{\rm eddy})$. MRI turbulence suggests that $t_{\rm eddy}\sim \Omega^{-1}$ and $\ell_{\rm eddy}\sim \sqrt{\alpha}H$ \citep[e.g.][]{zhu14}. Therefore, comparing
{\bc the} ratio of thermal time at the inertial scale to inertial time we find:
\begin{equation}
\frac{\tau_r}{t_{\rm eddy}}=A\frac{\alpha C_p\rho^2\kappa H^2\Omega}{12acT^3}
\end{equation}
where since $\rho$, $T$ \& $\kappa$ are slowly varying with radius (changing on a scale of $R$) and $\ell_{\rm eddy} < H \ll R$ then to lowest order in $\ell$ we can neglect the difference between the density, temperature and opacity between the fluid parcel and background to calculate $\tau_r$. Now assuming local thermal equilibrium  the temperature profile for a steady-alpha disc is:
\begin{equation}
T^4=\frac{27}{128\sigma}\kappa\Sigma^2\Omega\alpha c_s^2
\end{equation}
and remembering $\Sigma\sim \rho H$ then we can re-write the ratio $\tau_r/t_{\rm eddy}$ as:
\begin{equation}
\frac{\tau_r}{t_{\rm eddy}}\approx\frac{A}{10}
\end{equation}
for $\gamma=7/5$. This suggests that the integral scale of MRI turbulence is able to move fluid parcels adiabatically through the disc. Allowing for the diffusion of heat. However, since the eddy returns before the fluid parcel is able to radiate all its extra heat in to the surroundings then the integral scale will not be fully efficient at diffusing heat. In fact each overturn will be able to approximately transfer the fraction $t_r/t_{\rm eddy}$ of the heat to the surroundings. Thus, we may write $D^R_{\rm th}$ dimensionally as:
\begin{equation}
D^R_{th}=\langle u_R^2 \rangle \tau_r
\end{equation}
provided $\tau_r < t_{\rm eddy}$. Using the result for MRI turbulence that $\langle u_R^2 \rangle \sim \alpha c_s^2$ and $t_{\rm eddy} \sim \Omega^{-1}$ \citep[e.g.][]{zhu14} this suggest that a reasonable estimate of the Prandtl number is:
\begin{equation}
P_r=\frac{t_{\rm eddy}}{\tau_r}\approx\frac{10}{A}
\end{equation}
Indicating that turbulent diffusion of heat is likely to be extremely important in the properties of accretion bursts. We note we have used a steady disc approximation to evaluate the ratio of $\tau_r/t$; however, thermal diffusion is most important when the disc out of thermal equilibrium. Therefore, we suspect the Prandtl number may not be constant during the evolution. In particular, just prior to a burst the disc is cooler than one would expect from the steady disc approximation, so $t_{\rm eddy}/\tau_r$ would be smaller than our approximation suggests, decreasing the Prandtl number and increase the transport of heat due to turbulence. Alternatively as the burst comes to an end the steady disc approximation will overestimate the disc temperature and consequence the Prandtl number will be larger than our steady disc estimate suggest. This scenario would result in bursts being trigger quicker and lasting longer than simulations with constant Prandtl number indicate. The strong sensitivity of $\tau_r/t_{\rm eddy}$ on temperature for fixed surface density (in the range of a few hundred to thousand K $\kappa\sim  T$) results in $\tau_r/t_{\rm eddy}\sim 1/T^2$, indicating that a non-constant Prandtl number may have dynamically interesting consequences of the gravo-magnetic limit cycle.

\section{Conclusions}
We have investigated the role of thermal diffusion due to turbulence and radiation in the gravo-magnetic limit cycle in protoplanetary discs with a dead-zone using 1D numerical simulations. Our main conclusions are as follows:
\begin{enumerate}
\item Thermal diffusion strongly affects the time-scale of the limit cycle, where its conclusion decreases the burst length and increases the burst frequency. 
\item Thermal diffusion increases the range of accretion rates over which a limit cycle is obtained rather then steady accretion, with a limit cycle found for the range of infall rates from $10^{-5}-10^{-8}$ \msunyr for high Prandtl numbers and for small Prandtl numbers we found non-steady accretion above $\gtrsim 10^{-8}$~\msunyr upto the maximum simulated infall rate of $10^{-4}$~\msunyr . 
\item For Prandtl numbers $\lesssim 25$ we find turbulent diffusion is more important than radiative diffusion. 
\item For Prandtl numbers of unity the burst length is significantly shorter than previously simulated values with burst lengths $\sim 10^{3}$~years.
\item The duty-cycle is generally insensitive to the Prandtl number accept at low accretion rates (a few $10^{-8}$~\msunyr) where turbulent diffusion decreases the duty-cycle. 
\item An order of magnitude analysis suggests that Prandtl numbers of order 10 is a reasonable estimate for accretion discs, although it is unlikely to be constant and will depend on the thermodynamic state of the disc. 
\end{enumerate}

Since we have shown that if turbulent diffusion of heat is strong ($Pr\lesssim 25$) then it has important consequences for the properties of accretion bursts due to the gravo-magnetic limit cycle. This result should motivate further work to estimate the Prandtl number for MRI turbulence. Our very simple estimates indicate that the Prandtl number does indeed lie in the range where turbulent transport of heat is important; however, our calculation assumed local thermal equilibrium, and as such only numerical simulations that include the relevant physics can properly assess the role of turbulent diffusion of heat in accretion bursts driven in protoplanetary discs that contain dead zones.

\section*{Acknowledgments}
We thank the referee for a helpful report that improved this manuscript. 
We are grateful to Shane Davis, Emmanuel Jacquet and Kristen Menou for useful discussions. The numerical calculations were performed on the
Sunnyvale cluster at CITA, which is funded by the Canada Foundation for Innovation. PJA acknowledges support from NASA 
under grants NNX13AI58G and NNX14AB42G, from the NSF under award AST~1313021, and from grant HST-AR-12814 awarded by the Space Telescope Science Institute, which is operated by the Association of Universities for Research in Astronomy, Inc., for NASA, under contact NAS 5-26555

\end{document}